\documentclass[]{aa}
\usepackage{graphicx}
\usepackage{color}
\usepackage{txfonts}

\begin{document}

\title{Magnetohydrodynamic waves in solar partially ionized plasmas: two-fluid approach}

\author{Zaqarashvili, T.V.\inst{1,2}, Khodachenko, M.K. \inst{1} and Rucker, H.O. \inst{1}
}

 \institute{ Space Research Institute, Austrian Academy of Sciences, Schmiedlstrasse 6, 8042 Graz, Austria\\
             \email{[teimuraz.zaqarashvili;maxim.khodachenko;rucker]@oeaw.ac.at}
                               \and
            Abastumani Astrophysical Observatory at Ilia State University, Kazbegi ave. 2a, Tbilisi, Georgia\\
}

\date{Received / Accepted }

\abstract{Partially ionized plasma is usually described by single-fluid approach, where the ion-neutral collision effects are expressed by Cowling conductivity in the induction equation. However, the single-fluid approach is not valid for the time-scales less than ion-neutral collision time. For these time-scales the two-fluid description is better approximation.}{To derive the dynamics of magnetohydrodynamic waves in two-fluid partially ionized plasmas and to compare the results with those obtained under single-fluid description.}{Two-fluid magnetohydrodynamic equations are used, where ion-electron plasma and neutral particles are considered as separate fluids. Dispersion relations of linear magnetohydrodynamic waves are derived for simplest case of homogeneous medium. Frequencies and damping rates of waves are obtained for different parameters of background plasma.}{We found that two- and single-fluid descriptions give similar results for low frequency waves. However, the dynamics of MHD waves in two-fluid approach is significantly changed when the wave frequency becomes comparable or higher than ion-neutral collision frequency. Alfv\'en and fast magneto-acoustic waves attain their maximum damping rate at particular frequencies (for example, the peak frequency equals 2.5 ion-neutral collision frequency for 50 $\%$ of neutral Hydrogen) in wave spectrum. The damping rates are reduced for higher frequency waves. The new mode of slow magneto-acoustic wave appears for higher frequency branch, which is connected to neutral hydrogen fluid. }{The single-fluid approach perfectly deals with slow processes in partially ionized plasmas, but fails for time-scales smaller than ion-neutral collision time. Therefore, two-fluid approximation should be used for the description of relatively fast processes. Some results of single-fluid description, for example the damping of high-frequency Alfv\'en waves in the solar chromosphere due to ion-neutral collisions, should be revised in future. }

\keywords{Sun: atmosphere -- Sun: oscillations}

\titlerunning{MHD waves in partially ionized plasma}

\authorrunning{Zaqarashvili et al.}

\maketitle

\section{Introduction}

Astrophysical plasmas often are partially ionized. Neutral atoms may change the plasma dynamics due to collisions with charged particles. The ion-neutral collisions may lead to different new phenomena in plasma, for example the damping of magnetohydrodynamic (MHD) waves (Khodachenko el al. \cite{Khodachenko2004}, Forteza et al. \cite{Forteza2007}). Solar photosphere, chromosphere and prominences contain significant amount of neutral atoms, therefore the complete description of plasma processes requires the consideration of partial ionization effects.

Braginskii ({\cite{Braginskii1965}}) gave the basic principles of transport processes in plasma including the effects of partial ionization. Since this review, numerous papers addressed the problem of partial ionization in the different regions of solar atmosphere. Khodachenko and Zaitsev (\cite{Khodachenko2002}) studied the formation of magnetic flux tube in a converging flow of solar photosphere, while Vranjes et al. (\cite{Vranjes2008}) studied the Alfv\'en waves in weakly ionized photospheric plasma. Leake and Arber (\cite{Leake2005}) and Arber et al. ({\cite{Arber2007}}) studied the effect of partially ionized plasma on emerging magnetic flux tubes and concluded that the chromospheric neutrals may transform the magnetic tube into force-free configuration. Haerendel (\cite{Haerendel1992}), De Pontieu and Haerendel (\cite{De Pontieu1998}), James and Erd\'elyi (\cite{James2002}), James et al. (\cite{James2004}) considered the damping of Alfv\'en waves due to ion-neutral collision as a mechanism of spicule formation. Khodachenko el al. (\cite{Khodachenko2004}) and Leake et al. (\cite{Leake2006}) studied the importance of ion-neutral collisions in damping of MHD waves in the chromosphere and prominences. Forteza et al. (\cite{Forteza2007,Forteza2008}), Soler et al. (\cite{Soler2009a, Soler2009b,Soler2010}) and Carbonell et al. (\cite{Carbonell2010}) studied the damping of MHD waves in partially ionized prominence plasma with and without plasma flow.

All these papers considered the single-fluid MHD approach, when inertial terms in the momentum equation of relative velocity between ions and neutrals are neglected. The partially ionized plasma effects are described by generalized Ohm's law with Cowling conductivity, which leads to the modified  induction equation (Khodachenko el al. \cite{Khodachenko2004}). Ambipolar diffusion is more pronounced during the transverse motion of plasma with regards to magnetic field, therefore the Alfv\'en and fast magneto-acoustic wave are more efficiently damped. The slow magneto-acoustic waves are weakly damped in the low plasma beta case. Moreover, Forteza et al. (\cite{Forteza2007}) found that the damping rate of slow magneto-acoustic waves derived through normal mode analysis is different from that estimated by Braginskii ({\cite{Braginskii1965}}). The problem of discrepancy between normal mode analysis (Forteza et al. \cite{Forteza2007}) and energy consideration (Braginskii {\cite{Braginskii1965}}) is still an open question and the present study attempts to shed light on it.

The single-fluid approach has been shown to be valid for the time-scales which are larger than ion-neutral collision time. However, the approximation fails for the shorter time scales, therefore the two-fluid approximation, which means the treatment of ion-electron and neutral gases as separate fluids, should be considered. The two-fluid approximation is valid for the time-scales larger than ion-electron collision time, which is significantly short due to Coulomb collision between ions and electrons.

In this paper, we study MHD waves in two-fluid partially ionized plasma. The particular attention is paid to the wave damping due to ion-neutral
collisions and comparison between the wave dynamics in single and two-fluid approximations. We derive the two-fluid MHD equations from initial
three-fluid equations and solve the linearized equations in the simplest case of a homogeneous plasma.

\section{Main equations}

We aim to study partially ionized plasma, which consists in
electrons, ions and neutral atoms. We suppose that each sort of
spaces has Maxwell velocity distribution, therefore they can be
described as separate fluids. Below we first write the equations in three-fluid description and then perform consequent transition to two-fluid and single-fluid approaches.

\subsection{Three-fluid equations}

The fluid equations for each spaces can be derived from Boltzmann
kinetic equations and they have the following forms (Braginskii
\cite{Braginskii1965}, Goedbloed \& Poedts \cite{Goedbloed2004})

\begin{equation}\label{ne3}
{{\partial n_e}\over {\partial t}}+\nabla \cdot (n_e \vec V_e)=0,
\end{equation}
\begin{equation}\label{ni3}
{{\partial n_i}\over {\partial t}}+\nabla \cdot (n_i \vec V_i)=0,
\end{equation}
\begin{equation}\label{nn3}
{{\partial n_n}\over {\partial t}}+\nabla \cdot (n_n \vec V_n)=0,
\end{equation}
\begin{equation}\label{Ve3}
m_en_e\left ({{\partial \vec V_e}\over {\partial t}}+({\vec
V_e}\cdot \nabla)\vec V_e\right )=-\nabla p_{e}-\nabla \cdot \pi_e
-en_e\left (\vec E +{1\over c}\vec V_e \times \vec B \right )+\vec
R_e,
\end{equation}
\begin{equation}\label{Vi3}
m_in_i\left ({{\partial \vec V_i}\over {\partial t}}+({\vec
V_i}\cdot \nabla)\vec V_i\right )=-\nabla p_{i}-\nabla \cdot
\pi_i+Zen_i\left (\vec E +{1\over c}\vec V_i \times \vec B \right
)+\vec R_i,
\end{equation}
\begin{equation}\label{Vn3}
m_nn_n\left ({{\partial \vec V_n}\over {\partial t}}+({\vec
V_n}\cdot \nabla)\vec V_n\right )=-\nabla p_{n}-\nabla \cdot
\pi_n+\vec R_n,
\end{equation}
\begin{equation}\label{Te3}
{3\over 2}n_e k \left ({{\partial T_e}\over {\partial t}}+({\vec
V_e}\cdot \nabla)T_e\right )+p_e\nabla \cdot \vec V_e+\pi_e : \nabla
\vec V_e=-\nabla \cdot \vec q_e+Q_e
\end{equation}
\begin{equation}\label{Ti3}
{3\over 2}n_i k \left ({{\partial T_i}\over {\partial t}} +({\vec
V_i}\cdot \nabla)T_i\right )+p_i\nabla \cdot \vec V_i+\pi_i : \nabla
\vec V_i=-\nabla \cdot \vec q_i+Q_i
\end{equation}
\begin{equation}\label{Tn3}
{3\over 2}n_n k \left ({{\partial T_n}\over {\partial t}} +({\vec
V_n}\cdot \nabla)T_n\right )+p_n\nabla \cdot \vec V_n+\pi_n : \nabla
\vec V_n=-\nabla \cdot \vec q_n+Q_n
\end{equation}
\begin{equation}\label{p3}
p_e=n_ekT_e,\,\,p_i=n_ikT_i,\,\,p_n=n_nkT_n,
\end{equation}
where $m_a$, $n_a$, $p_a$, $T_a$, $\vec V_a$ are the mass, the
density, the pressure, the temperature and the velocity of particles
$a$, $\vec E$ is the electric field, $\vec B $ is the magnetic field
strength, $\vec q_a$ is the heat flux density of particles $a$,
$\vec R_a$ is the change of impulse of particles $a$ due to
collisions with other sort of particles, $Q_a$ is the heat
production due to collisions of particles $a$ with other sort of
particles, $\pi_a$ is the off-diagonal pressure tensor of particles
$a$, $e=4.8\times 10^{-10}$ statcoul is the electron charge,
$c=2.9979\times 10^{10}$ cm s$^{-1}$ is the speed of light and
$k=1.38\times 10^{-16}$ erg K$^{-1}$ is the Boltzmann constant. The
double dot indicates that a double sum over the Cartesian components
is to be taken. Plasma is supposed to be quasi-neutral, which means
$n_e=Zn_i$. In what follows we consider hydrogen ions and hydrogen neutral atoms which imply
$Z=1$. The description of the system is completed by Maxwell
equations which have the forms (without displacement current)
\begin{equation}\label{e}
\nabla \times \vec E=-{1\over c}{{\partial \vec B}\over {\partial t}},
\end{equation}
\begin{equation}\label{B}
\nabla \times \vec B={{4 \pi} \over c}{\vec j},
\end{equation}
where
\begin{equation}\label{j2}
\vec j=-en_e(\vec V_e-\vec V_i)=-en_e\vec u
\end{equation}
is the current density.

In the case of Maxwell distribution in each sort of particles, $\vec
R_a$ and $Q_a$ are expressed as the following (Braginskii \cite{Braginskii1965}):
\begin{equation}\label{Re}
\vec R_e=-\alpha_{ei}(\vec V_e-\vec V_i)-\alpha_{en}(\vec V_e-\vec V_n),
\end{equation}
\begin{equation}\label{Ri}
\vec R_i=-\alpha_{ie}(\vec V_i-\vec V_e)-\alpha_{in}(\vec V_i-\vec V_n),
\end{equation}
\begin{equation}\label{Rn}
\vec R_n=-\alpha_{ne}(\vec V_n-\vec V_e)-\alpha_{ni}(\vec V_n-\vec V_i),
\end{equation}
\begin{equation}\label{Qe}
Q_e=\alpha_{ei}(\vec V_e-\vec V_i)\vec V_e+\alpha_{en}(\vec V_e-\vec V_n)\vec V_e,
\end{equation}
\begin{equation}\label{Qi}
Q_i=\alpha_{ie}(\vec V_i-\vec V_e)\vec V_i+\alpha_{in}(\vec V_i-\vec V_n)\vec V_i,
\end{equation}
\begin{equation}\label{Qn}
Q_n=\alpha_{ne}(\vec V_n-\vec V_e)\vec V_n+\alpha_{ni}(\vec V_n-\vec V_i)\vec V_n,
\end{equation}
where $\alpha_{ab}=\alpha_{ba}$ are coefficients of friction between particles $a$ and $b$.

%

For time scales longer than ion-electron collision time, the
electron and ion gases can be considered as a single fluid. This
significantly simplifies the equations taking into account the
smallness of electron mass with regards to the masses of ion and
neutral atoms. Then the three-fluid description can be changed by
two-fluid description, where one component is ion-electron gas and
the second component is the gas of neutral atoms.

\subsection{Two-fluid equations}

Summing of Eqs. (\ref{Ve3}) and (\ref{Vi3}), Eqs. (\ref{Te3})
and (\ref{Ti3}) and first two equations of Eq. (\ref{p3}), we obtain
(after neglecting the electron inertia and the viscosity effect expressed by
off-diagonal pressure tensor $\pi_a$)
\begin{equation}\label{ni2}
{{\partial n_i}\over {\partial t}}+\nabla \cdot (n_i \vec V_i)=0,
\end{equation}
\begin{equation}\label{nn2}
{{\partial n_n}\over {\partial t}}+\nabla \cdot (n_n \vec V_n)=0,
\end{equation}
$$
m_in_i\left ({{\partial \vec V_i}\over {\partial t}}+({\vec
V_i}\cdot \nabla)\vec V_i\right ) =-\nabla p_{ie}+{1\over c}\vec j
\times \vec B+{{\alpha_{en}}\over {e n_e}}\vec j -
$$
\begin{equation}\label{vi2}
(\alpha_{in}+\alpha_{en})(\vec V_i -\vec V_n),
\end{equation}
\begin{equation}\label{vn2}
m_nn_n\left ({{\partial \vec V_n}\over {\partial t}}+({\vec
V_n}\cdot \nabla)\vec V_n\right ) =-\nabla p_n-{{\alpha_{en}}\over
{e n_e}}\vec j + (\alpha_{in}+\alpha_{en})(\vec V_i -\vec V_n),
\end{equation}
$$
{{\partial p_{ie}}\over {\partial t}}+ ({\vec V_i}\cdot
\nabla)p_{ie} + \gamma p_{ie} \nabla \cdot \vec V_i =(\gamma -1)
{{\alpha_{ei}}\over {e^2 n^2_e}}j^2 +(\gamma -1)\alpha_{in} (\vec
V_i -\vec V_n) \cdot \vec V_i+
$$
\begin{equation}\label{p2}
(\gamma -1)\alpha_{en}(\vec V_e -\vec
V_n) \cdot \vec V_e+{{({\vec j}\cdot \nabla)p_{e}}\over {e n_e}}+\gamma p_{e}\nabla
\cdot {{\vec j}\over {e n_e}} -
(\gamma -1)\nabla \cdot ({\vec
q_i}+{\vec q_e}),
\end{equation}
$$
{{\partial p_{n}}\over {\partial t}}+ ({\vec V_n}\cdot \nabla)p_{n}+
\gamma p_{n} \nabla \cdot \vec V_n=-(\gamma -1)\alpha_{in} (\vec V_i
-\vec V_n) \cdot \vec V_n+
$$
\begin{equation}\label{pn2}
(\gamma -1)\alpha_{en} (\vec V_n -\vec
V_e) \cdot \vec V_n -(\gamma -1)\nabla \cdot {\vec q_n},
\end{equation}
where $p_{ie}=p_i+p_e$ is the pressure of ion-electron gas and
$\gamma=C_p/C_v=5/3$ is the ratio of specific heats.

The Ohm's law is obtained from the electron equation (Eq. \ref{Ve3})
after neglecting the electron inertia (i.e. the left-hand side
terms) and it has the form
\begin{equation}\label{om2}
\vec E + {1\over c}\vec V_i \times \vec B+ {{1}\over {e n_e}}\nabla
p_e ={{\alpha_{ei}+\alpha_{en}}\over {e^2 n^2_e}}\vec
j-{{\alpha_{en}}\over {e n_e}}(\vec V_i -\vec V_n) +{1\over
{cen_e}}\vec j\times \vec B.
\end{equation}

Maxwell equation (Eq. \ref{e}) and Ohm's law (Eq. \ref{om2}) lead to
the induction equation
$$
{{\partial \vec B}\over {\partial t}}={\nabla \times}(\vec V_i
\times \vec B)+{\nabla \times}\left ({{c \nabla p_e}\over {e
n_e}}\right )-{\nabla \times}\left (\eta \nabla \times \vec B\right
) - {\nabla \times}\left ({{\vec j\times \vec B}\over {en_e}}\right
)+
$$
\begin{equation}\label{B2}
{\nabla \times}\left ({{c \alpha_{en}(\vec V_i -\vec V_n)}\over {e
n_e}}\right ),
\end{equation}
where
\begin{equation}\label{eta}
\eta={{c^2}\over {4\pi \sigma}}={{c^2(\alpha_{ei}+\alpha_{en})}\over {4\pi e^2 n^2_e}}
\end{equation}
is the coefficient of magnetic diffusion.

The coefficient of friction between ions and neutrals (in the case of same temperature) is calculated as (Braginskii \cite{Braginskii1965})
\begin{equation}\label{alfev-disp1}
{\alpha_{in}}= n_i n_n m_{in}\sigma_{in}{4\over 3} \sqrt{{{8kT}\over {\pi m_{in}}}}  .
\end{equation}
where $m_{in}=m_i m_n /(m_i+m_n)$ is reduced mass and $\sigma_{in}=\pi (r_i+r_n)^2=4\pi
r^2_i$ is the collision cross section between ions and neutrals.

Collision frequency between ions and neutrals is then
$$
\nu_{in}={{\alpha_{in}}\over
{m_i n_{i}+m_n n_{n}}}={{16\pi r^2_i}\over 3} {{n_i n_n m_{in}}\over
{m_i n_{i}+m_n n_{n}}} \sqrt{{{8kT}\over {\pi m_{in}}}} =
$$
\begin{equation}\label{collision}
{{32\pi r^2_i}\over {3\sqrt{\pi}}} {{n_i n_n}\over
{n_{i}+n_{n}}} \sqrt{{{kT}\over {m_{i}}}}\approx 5\times 10^{-12}{{n_i n_n}\over
{n_{i}+n_{n}}}\sqrt{T} \,\, s^{-1},
\end{equation}
where the atomic cross section  $\pi r^2_i=8.7974\times 10^{-17}$
cm$^{2}$ is used and $T$ is normalised by 1 $K$. The chromospheric temperature of $10^4$ K
and hydrogen ion and neutral number densities of $2.3\times 10^{10}$ cm$^{-3}$
and $1.2\times 10^{10}$ cm$^{-3}$ (Fontenla et al. \cite{Fontenla1990}, model FAL-3) give the collision frequency as 4 s$^{-1}$.

For time scales longer than ion-neutral collision time ($1/\nu_{in}$), the system
can be considered as a single fluid (the full equations of
single-fluid MHD including neutral hydrogen are presented in
AppendixA). However, when the time scales are near or shorter than
ion-neutral collision time, then the single-fluid description is not
valid and the two-fluid equations should be considered.
In what follows we study the linear MHD waves in two-fluid
description.

\section{Linear MHD waves}

We consider the simplest case of static and homogeneous plasma with
homogeneous unperturbed magnetic field. Then the linearized
two-fluid equations followed from Eqs. (\ref{ni2})-(\ref{pn2}) and (\ref{B2}) are (neglecting the Hall term and the
collision between neutrals and electrons i.e. $\alpha_{en}\ll \alpha_{in}$):
\begin{equation}\label{ni2l}
{{\partial \rho^{\prime}_i}\over {\partial t}}+\rho_{i0} \nabla \cdot \vec v_i=0,
\end{equation}
\begin{equation}\label{nn2l}
{{\partial \rho^{\prime}_n}\over {\partial t}}+\rho_{n0} \nabla \cdot \vec v_n=0,
\end{equation}
$$
\rho_{i0}{{\partial \vec v_i}\over {\partial t}}=-\nabla
p^{\prime}_{ie}-{1\over {4 \pi}}\nabla (\vec B_0 \cdot \vec b)
+{1\over {4 \pi}}(\vec B_0 \cdot \nabla) \vec b + {{\alpha_{en}
c}\over {4 \pi e n_e}} \nabla \times \vec b -
$$
\begin{equation}\label{ui2}
{{\alpha_{in}}}(\vec
v_i -\vec v_n),
\end{equation}
\begin{equation}\label{un2}
\rho_{n0}{{\partial \vec v_n}\over {\partial t}}=-\nabla
p^{\prime}_n - {{\alpha_{en} c}\over {4 \pi e n_e}} \nabla \times
\vec b + {{\alpha_{in}}} (\vec v_i -\vec v_n),
\end{equation}
\begin{equation}\label{b2}
{{\partial \vec b}\over {\partial t}}=(\vec B_0 \cdot \nabla) \vec
v_i- \vec B_0  \nabla \cdot \vec v_i+\eta \nabla^2 \vec b+{{c
\alpha_{en}}\over {e n_e}}{\nabla \times} (\vec v_i -\vec v_n),
\end{equation}
\begin{equation}\label{p2l}
{{\partial p^{\prime}_{ie}}\over {\partial t}}+\gamma p_{ie} \nabla \cdot \vec v_i=0,
\end{equation}
\begin{equation}\label{pn2l}
{{\partial p^{\prime}_{n}}\over {\partial t}}+\gamma p_{n} \nabla \cdot \vec v_n=0,
\end{equation}
where $\rho^{\prime}_i$ ($\rho^{\prime}_n$) are perturbations of ion (neutral) density, $\vec v_i$ ($\vec v_n$) are the perturbations of ion (neutral) velocity, $p^{\prime}_{ie}$ ($p^{\prime}_{n}$) are the perturbations of ion-electron (neutral) gas pressure, $\vec b$ is the perturbation of the magnetic field, and $\rho_{i0}=m_in_{i0}, \rho_{n0}=m_nn_{n0}, p_{ie}, p_{n}, \vec
B_0$ are
their unperturbed values, respectively. Eqs. (\ref{ni2l})-(\ref{nn2l}) and Eqs.
(\ref{p2l})-(\ref{pn2l}) lead to the expressions
\begin{equation}\label{ad}
p^{\prime}_{ie}= c^2_{si}\rho^{\prime}_i,\,\, p^{\prime}_{n}= c^2_{sn}\rho^{\prime}_n,
\end{equation}
where $c_{si}=\sqrt{\gamma p_{ie}/\rho_{i0}}$ and
$c_{sn}=\sqrt{\gamma p_n/\rho_{n0}}$ are sound speeds of
ion-electron and neutral gases, respectively.

Below we consider the unperturbed magnetic field, $B_z$, directed
along the $z$ axis and wave propagation in $xz$ plane i.e. $\partial
/\partial y=0$. Then Eqs. (\ref{ni2l})-(\ref{pn2l})
can be split into Alfv\'en and magneto-acoustic waves.

\subsection{Alfv\'en waves}

Let us assume the Alfv\'en waves polarized along $y$ axis. We
intend to study the damping of Alfv\'en waves due to collision
between ions and neutrals. Therefore, we neglect the magnetic
diffusion for simplicity. Then,  Eqs. (\ref{ni2l})-(\ref{pn2l}) give
\begin{equation}\label{alfen-viy}
{{\partial v_{iy}}\over {\partial t}}={B_z\over {4 \pi
\rho_{i0}}}{{\partial b_{y}}\over {\partial z}} -
{{\alpha_{in}}\over \rho_{i0}} (v_{iy} -v_{ny}),
\end{equation}
\begin{equation}\label{alfven-vny}
{{\partial v_{ny}}\over {\partial t}}={{\alpha_{in}}\over \rho_{n0}}
(v_{iy} -v_{ny}),
\end{equation}
\begin{equation}\label{alfven-by}
{{\partial b_{y}}\over {\partial t}}=B_z{{\partial v_{iy}}\over {\partial z}}.
\end{equation}

\begin{figure}[t]
\vspace*{1mm}
\begin{center}
\includegraphics[width=9.5cm]{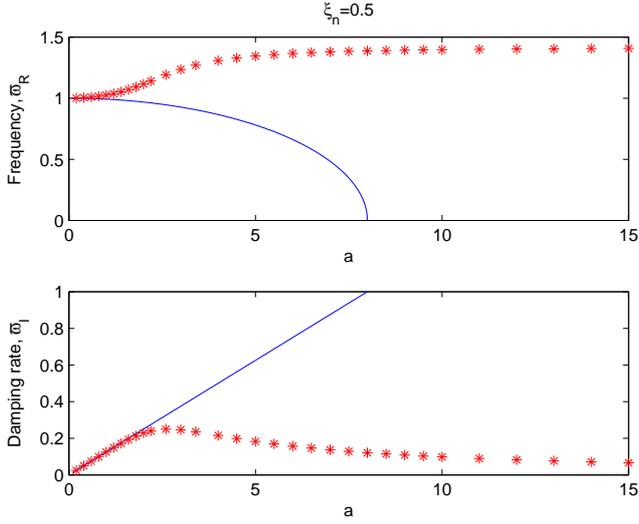}
\end{center}
\caption{Real ($\varpi_R$) and imaginary ($\varpi_I$) parts of
Alfv\'en wave frequency, ${\varpi}$, vs $a$ (where $a={{k_z
v_A}/ {\nu_{in}}}$). The blue line corresponds to the solution of
single-fluid dispersion relation, i.e. Eq. (\ref{alfev-disp1}) and
red asterisks are the solutions of two-fluid dispersion relation,
Eq. (\ref{alfev-disp2}). The values are calculated for 50\% of
neutral hydrogen, $\xi_n$=0.5.}
\end{figure}

\begin{figure}
\begin{center}
\includegraphics[width=9.5cm]{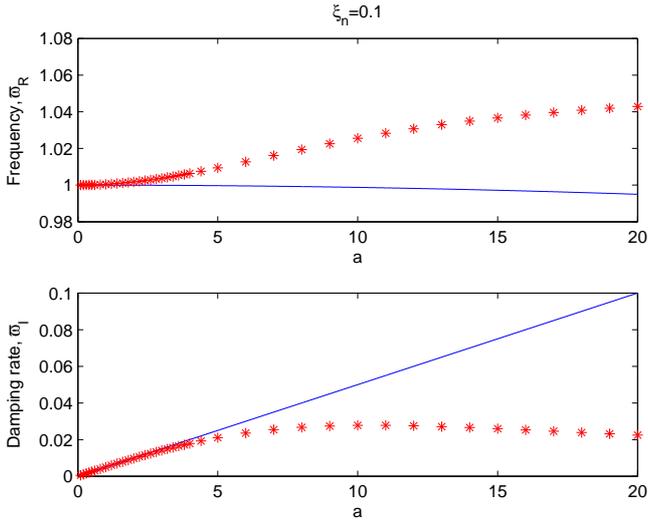}
\end{center}
\caption{Same as on Fig.1, but for 10\% of neutral hydrogen, $\xi_n$=0.1.}
\end{figure}

Fourier analyses assuming disturbances to be proportional to $exp[i(k_z z -\omega t)]$ give the dispersion relation

\begin{equation}\label{alfev-disp2}
a \xi_i \xi_n {\varpi}^3 + i{\varpi}^2 -
a \xi_n {\varpi}-i =0,
\end{equation}
where
$$
{\varpi}={{\omega}\over {k_zv_A}},\,\,a={{k_z v_A}\over
{\nu_{in}}},\,\, \xi_i={\rho_i\over \rho_0}, \,\, \xi_n={\rho_n\over
\rho_0}, \,\,  v_A= {B_z\over {\sqrt{4 \pi \rho_{0}}}}, \,\,
$$
\begin{equation}\label{alfven-speed}
\nu_{in}={\alpha_{in}\over \rho_0},\,\,\rho_0=\rho_{i0}+\rho_{n0}.
\end{equation}

The same dispersion relation can be obtained from linear single-fluid
equations retaining the inertial term in Eq. (\ref{w1}). The dispersion relation of the Alfv\'en waves in linear single-fluid equations without the inertial term can be easily derived as
\begin{equation}\label{alfev-disp1}
{\varpi}^2+i a \xi^2_n {\varpi} -1=0.
\end{equation}

The solution of Eq. (\ref{alfev-disp1}) is
\begin{equation}\label{sol-alfev}
{\varpi}={{-i a \xi^2_n \pm \sqrt{-a^2 \xi^4_n+4}}\over 2},
\end{equation}
which for $a \xi^2_n<2$ gives the damping rate
\begin{equation}\label{sol-alfev}
2{\omega_i}={{\xi^2_n B^2_z}\over {4\pi \alpha_{in}}}k^2_z
\end{equation}
in full coincidence with Braginskii (\cite{Braginskii1965}). On the
other hand, the condition $a \xi^2_n>2$ in Eq. (\ref{sol-alfev})
retains only imaginary part, which gives the cut-off wave number
\begin{equation}\label{cut-off}
k_{c}={{2 \nu_{in}}\over {\xi^2_n v_A}}.
\end{equation}
The value of cut-off wave number has been obtained recently by
Barc\'elo et al. (\cite{Barcelo2010}). Hence, the waves with
higher wave number than ${k_{c}}$ are evanescent. However, it might be incomplete conclusion as the complete treatment requires inclusion of inertial terms, and therefore dealing with Eq. (\ref{alfev-disp2}) instead of Eq. (\ref{alfev-disp1}). The first term in Eq. (\ref{alfev-disp2})
is important for the high frequency part of wave spectrum and could
not be neglected. We demonstrate it by solutions of Eqs.
(\ref{alfev-disp1}) and (\ref{alfev-disp2}).

\begin{figure}[t]
\vspace*{1mm}
\begin{center}
\includegraphics[width=9.5cm]{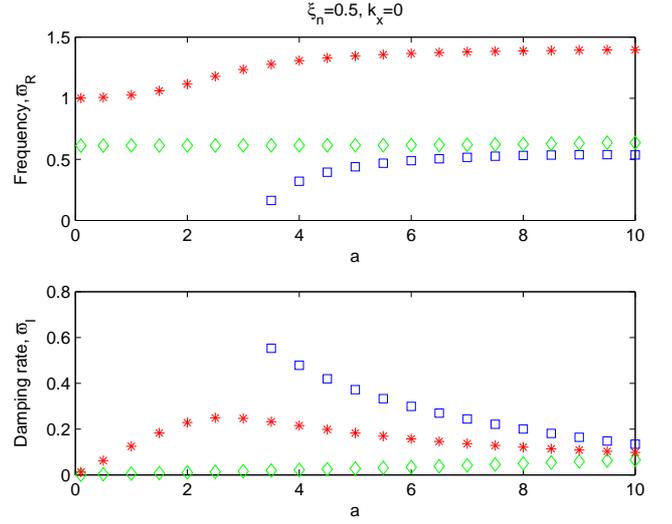}
\end{center}
\caption{Frequency and damping rate of different wave modes in two-fluid MHD vs $a$ (where $a={{k v_A}/ {\nu_{in}}}$). Frequencies and damping rates are normalized by $kv_A$. Red asterisks correspond to fast magneto-acoustic mode and green diamonds correspond to usual slow magneto-acoustic mode. The mode with the blue squares is the new sort of slow magneto-acoustic wave ("neutral" slow mode), which arises for larger wave numbers. This mode has only imaginary frequency for small wave numbers, which is not shown in this figure. Frequencies and damping rates are calculated for the waves propagating along the magnetic field. The neutral hydrogen is taken to be 50\% ($\xi_n$=0.5). Here we consider $c_{sn}/v_A=0.5$. }
\end{figure}

Fig. 1 displays the solutions of single-fluid (Eq.
\ref{alfev-disp1}, blue lines) and two-fluid (Eq. \ref{alfev-disp2},
red asterisks) dispersion relations for $\xi_n=0.5$. We see that the
frequencies and damping rates of Alfv\'en waves are same in
single-fluid and two-fluid approaches for low-frequency branch of
spectrum (small $a$). But the behavior is dramatically changed
when the wave frequency becomes comparable or higher than the
ion-neutral collision frequency, $\nu_{in}$, i.e. for $a>1$. The
damping time linearly increases with $a$  and the wave frequency
becomes zero at some point in single-fluid case (blue lines). The
point where the wave frequency becomes zero corresponds to the
cut-off wave number ${k_{c}}$ of Barc\'elo et al. (\cite{Barcelo2010}). However, there is no cut-off wave
number in solutions of two-fluid dispersion relation (red
asterisks): Eq. (\ref{alfev-disp2}) always has a solution with a real part.
Therefore, {\it the occurrence of the cut-off wave number in
single-fluid description is the result of neglecting the
inertial terms in the momentum equation of relative velocity between ions and neutrals}. Therefore, Eq. (\ref{alfev-disp2}) is the
correct dispersion relation for the whole spectrum of waves. But the dispersion relation (\ref{alfev-disp1}) is still
good approximation for lower frequency part of spectrum. Another interesting point of two-fluid approach
is that the damping rate (i.e. $\omega_I$) attains its maximal value
at some wave-lengths for which $a \approx 2.5$. The damping rate
decreases for smaller and larger $a$. This means that the waves,
which have the frequency in the interval $\nu_{in} < \omega < 10 \,
\nu_{in}$, have stronger damping than other harmonics of spectrum.
This is totally different from single-fluid solutions, which show
the linear increase of damping rate with increasing wave number
(lower panel, blue line).

Fig. 2 displays the same solutions as on the Fig. 1, but for
$\xi_n=0.1$. The solutions have basically same properties as those
with $\xi_n=0.5$. However, the wave length with maximal damping rate
is now shifted to $a \approx 10$.

\subsection{Magneto-acoustic waves}

Now let us turn to magneto-acoustic waves. We consider the waves and
wave vectors polarized in $xz$ plane. Then Eqs.
(\ref{ni2l})-(\ref{pn2l}) are written as (magnetic diffusion is again
neglected)
\begin{equation}\label{ni2lc}
{{\partial \rho^{\prime}_i}\over {\partial t}}+\rho_{i0}\left
({{\partial v_{ix}}\over {\partial x}}+ {{\partial v_{iz}}\over
{\partial z}} \right )=0,
\end{equation}
\begin{equation}\label{nn2lc}
{{\partial \rho^{\prime}_n}\over {\partial t}}+\rho_{n0}\left
({{\partial v_{nx}}\over {\partial x}}+{{\partial v_{nz}}\over
{\partial z}} \right )=0,
\end{equation}
\begin{equation}\label{ui2xc}
{{\partial v_{ix}}\over {\partial t}}= -{1\over \rho_{i0}}{{\partial
p^{\prime}_{ie}}\over {\partial x}}-{B_z\over {4 \pi
\rho_{i0}}}{{\partial b_z}\over {\partial x}}+{B_z\over {4 \pi
\rho_{i0}}}{{\partial b_x}\over {\partial z}}-{\alpha_{in}\over
\rho_{i0}}(v_{ix}-v_{nx}),
\end{equation}
\begin{equation}\label{ui2zc}
{{\partial v_{iz}}\over {\partial t}}= -{1\over \rho_{i0}}{{\partial
p^{\prime}_{ie}}\over {\partial z}}-{\alpha_{in}\over
\rho_{i0}}(v_{iz}-v_{nz}),
\end{equation}
\begin{equation}\label{un2xc}
{{\partial v_{nx}}\over {\partial t}}= -{1\over \rho_{n0}}{{\partial
p^{\prime}_{n}}\over {\partial x}}+{\alpha_{in}\over
\rho_{n0}}(v_{ix}-v_{nx}),
\end{equation}
\begin{equation}\label{un2xc}
{{\partial v_{nz}}\over {\partial t}}= -{1\over \rho_{n0}}{{\partial
p^{\prime}_{n}}\over {\partial z}}+{\alpha_{in}\over
\rho_{n0}}(v_{iz}-v_{nz}),
\end{equation}
\begin{equation}\label{b2c}
{{\partial b_x}\over {\partial t}}= B_z {{\partial v_{ix}}\over {\partial z}},
\end{equation}
\begin{equation}\label{p2lc}
{{\partial p^{\prime}_{ie}}\over {\partial t}}+\gamma p_{ie}\left
({{\partial v_{ix}}\over {\partial x}}+ {{\partial v_{iz}}\over
{\partial z}} \right )=0,
\end{equation}
\begin{equation}\label{pn2lc}
{{\partial p^{\prime}_{n}}\over {\partial t}} +\gamma p_{n} \left
({{\partial v_{nx}}\over {\partial x}}+ {{\partial v_{nz}}\over
{\partial z}} \right )=0.
\end{equation}

\begin{figure}[t]
\vspace*{1mm}
\begin{center}
\includegraphics[width=9.5cm]{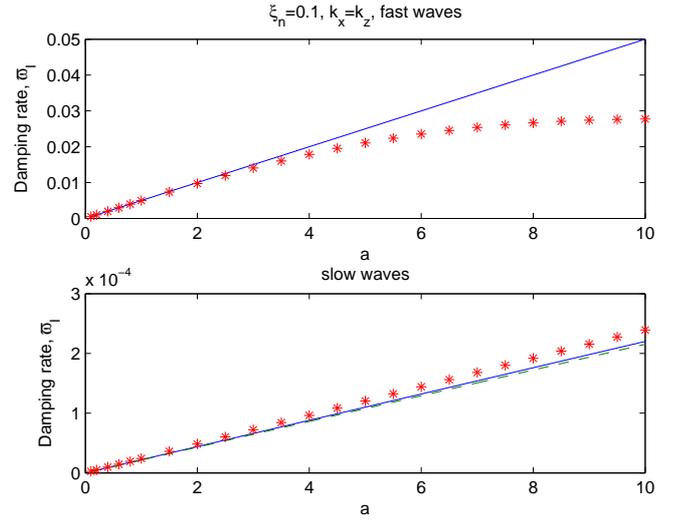}
\end{center}
\caption{Damping rate of fast (upper panel) and slow (lower panel) magneto-acoustic waves, i.e imaginary part of $\omega$ normalized by $kv_A$, vs $a$ (where $a={{k v_A}/ {\nu_{in}}}$). The blue solid lines corresponds to the solution of
single-fluid dispersion relation and dashed line corresponds to the slow magneto-acoustic damping rate of Braginskii (the expressions are used from   Forteza et al. \cite{Forteza2007}). Red asterisks are the solutions of two-fluid dispersion relation,
Eq. (\ref{disp-mhd}). The values are calculated for 10\% of
neutral hydrogen, $\xi_n$=0.1, and for $c_{sn}/v_A=0.1$. The damping rates are calculated for the waves propagate with 45$^{0}$ angle with regards to the magnetic field.   }
\end{figure}

Fourier analysis with $exp[i(k_x x+k_z z -\omega t)]$ and some algebra give the dispersion relation
$$
\nu^2_{in} \omega\left [\omega^4-k^2 (c^2_{si} \xi_i + c^2_{sn}
\xi_n + V^2_A) \omega^2+(c^2_{si}\xi_i + c^2_{sn} \xi_n) k^2 k^2_z
V^2_A \right ] -
$$
$$
\xi_i \xi^2_n \omega (\omega^2-c^2_{sn} k^2) \left
[\xi_i \omega^4-k^2 V^2_A \omega^2 +c^2_{si} k^2 (k^2_z V^2_A -
\xi_i \omega^2)\right ] -
$$
$$
i \nu_{in} \xi_n [(\xi_n-2) k^2 V^2_A \omega^4 +2 \xi_i
\omega^6 +c^2_{sn} k^2 \omega^2 \left (k^2 V^2_A +
(\xi^2_n-1)\omega^2 \right )+
$$
\begin{equation}\label{disp-mhd}
c^2_{si} \xi_i k^2 \left (2 k^2_z V^2_A
\omega^2 + (\xi_n-2) \omega^4 + c^2_{sn} k^2 (\omega^2-k^2_z V^2_A)
\right )]=0,
\end{equation}
where $k=\sqrt{k^2_x+k^2_z}$.

The dispersion relation  (\ref{disp-mhd}) is seventh order equation with $\omega$, therefore it has 7 different solutions. For smaller wave numbers (or smaller frequencies) four of the solutions represent the usual magneto-acoustic waves, while 3 other solutions are purely imaginary and are probably connected to the vortex modes (with $Re(\omega)=0$) damped due to ion neutral collisions. The vortex modes are solutions of fluid equations and they correspond to the fluid vorticity. The vortex modes have zero frequency in the ideal fluid, but may gain purely imaginary frequency if dissipative processes are evolved. The two vortex modes are transformed into oscillatory modes for shorter wavelengths (see the next paragraph). Then we have two fast magneto-acoustic modes, four slow magneto-acoustic modes and one vortex solution with purely imaginary part. In what follows, we consider that the temperatures of all three species are equal i.e. $T_i=T_e=T_n$, which gives $c^2_{si}=\gamma p_{ie}/\rho_{i0}=\gamma (p_i +p_e)/\rho_{i0}=\gamma k(T_i +T_e)/m_i= 2\gamma k T_n/m_n= 2 \gamma p_{n}/\rho_{n0}=2 c^2_{sn}$.

Fig. 3 displays all oscillatory solutions of two-fluid dispersion relation for $\xi_n=0.5$ (only the modes with positive frequencies are shown). The wave propagation is parallel to the magnetic field and we use $c_{sn}/v_A=0.5$, where $c_{sn}$ is the sound speed of neutral hydrogen. For smaller wave numbers, $k<3.5 \, \nu_{in}/v_A$, there are two usual magneto-acoustic modes, fast (red asterisks) and slow (green diamonds). However, for larger wave-numbers, $k>3.5 \, \nu_{in}/v_A$ one additional sort of slow magneto-acoustic mode with strong damping rate (blue squares) arises. The "neutral" slow mode is connected with neutral atoms. For higher frequency range, i.e. for those larger than ion-neutral collision frequency, neutral gas does not feel the ions, therefore it supports the propagation of additional oscillatory wave mode. This mode obviously disappears for lower frequency as the collisions couple ions and neutrals and they behave as a single fluid. In other words, for lower frequencies (or small wave numbers) this mode has zero real part, but non-zero imaginary part (not shown in the figure). Therefore, more correct statement is that the oscillatory mode transforms into non-oscillatory vortex mode for smaller wave numbers. The fast magneto-acoustic modes decouple from the slow waves for the parallel propagation and they have the same behavior as Alfv\'en waves. Therefore, the plot of fast magneto-acoustic waves is similar to that of Alfv\'en waves (see Fig.1).

\begin{figure}[t]
\vspace*{1mm}
\begin{center}
\includegraphics[width=9.5cm]{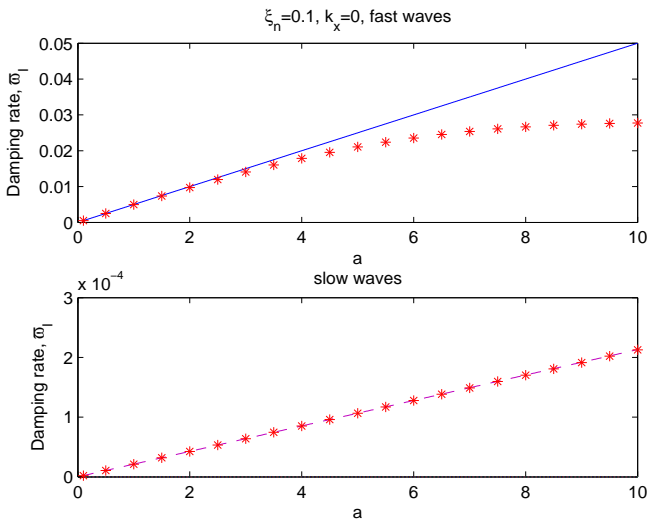}
\end{center}
\caption{The same as in Fig. 4 but for the parallel propagation, i.e. $k_x=0$.}
\end{figure}

It is useful to compare the solutions of two-fluid dispersion relation to those obtained in the single-fluid approach. The damping of fast and slow magneto-acoustic waves have been derived from the energy equation by Braginskii (\cite{Braginskii1965}), Khodachenko et al. (\cite{Khodachenko2004}), Khodachenko \& Rucker (\cite{Khodachenko2005}) and through normal mode analysis by Forteza et al. (\cite{Forteza2007}). Damping rates are same in both considerations for fast magneto-acoustic waves, but they disagree for slow magneto-acoustic waves (Forteza et al. \cite{Forteza2007}). Namely, slow magneto-acoustic waves have damping for purely parallel propagation in the case of Braginskii, while the damping is absent in the case of Forteza et al. Fig. 4 and 5 show the damping rates of fast and slow magneto-acoustic waves vs $a$ (i.e. $k$) for the propagation angles of $45^0$ and $0^0$, respectively. Red asterisks are the solutions of two-fluid dispersion relation-
Eq. (\ref{disp-mhd}). The blue solid lines correspond to the solutions of single-fluid dispersion relation from Forteza et al. (\cite{Forteza2007}).
The dashed line corresponds to the slow magneto-acoustic damping rate of Braginskii (\cite{Braginskii1965}). Here we use $c_{sn}/v_A=0.1$ so the  plasma $\beta$ is small enough. Neutral hydrogen concentration is taken to be 10$\%$. The fast magneto-acoustic waves have essentially same dynamics as the Alfv\'en waves. For the smaller wave numbers (or smaller frequencies) the two-fluid and single-fluid approaches give the same results, but for the larger wave numbers the damping rate is decreased in two-fluid description as in the case of Alfv\'en waves. On the other hand, the slow magneto-acoustic waves have similar damping rates in both approaches. There is a small discrepancy between damping rates for the waves propagating with $45^0$ degree about the magnetic field, but all the three cases (two-fluid waves, single-fluid waves and energy consideration) give similar results. The parallel propagation reveals an interesting result: two-fluid solutions are exactly the same as those obtained by Braginskii (the damping rate obtained by Forteza et al. \cite{Forteza2007} is zero for the parallel propagation). Therefore, the discrepancy between damping rates of slow magneto-acoustic waves obtained by Forteza et al. \cite{Forteza2007}) and Braginskii (\cite{Braginskii1965}) is again caused by neglecting the inertial terms in the momentum equation of relative velocity (Eq. \ref{w1}). Braginskii (\cite{Braginskii1965}) used energy equation to calculate the damping rate, therefore his solution agrees to that obtained in two-fluid approach.

\begin{figure}[t]
\vspace*{1mm}
\begin{center}
\includegraphics[width=9.5cm]{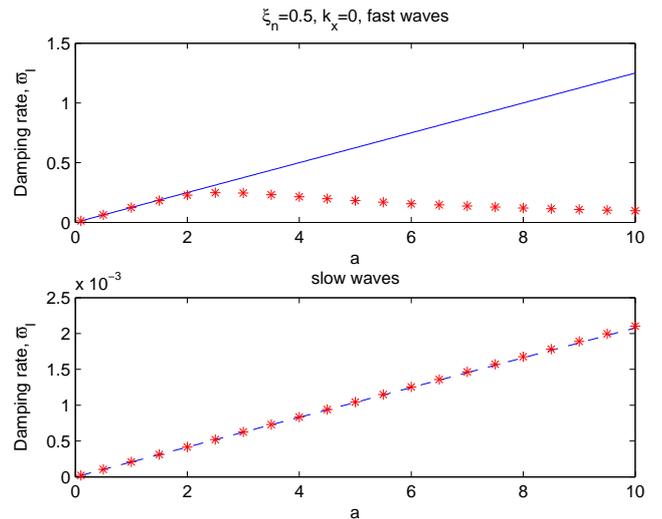}
\end{center}
\caption{Damping rate of fast (upper panel) and slow (lower panel) magneto-acoustic waves vs $a$ . Red asterisks are the solutions of two-fluid dispersion relation. Blue lines correspond to the solutions of Braginskii. The values are calculated for 50\% of
neutral hydrogen and $c_{sn}/v_A=0.1$. The damping rates are calculated for the waves propagate along the magnetic field. }
\end{figure}

Fig. 6 shows the comparison of damping rates in two-fluid approach and those obtained by Braginskii (\cite{Braginskii1965}) in the case of
$\xi_n=0.5$, $c_{sn}/v_A=0.1$ and parallel propagation. The fast magneto-acoustic waves have the same bihaviour as the Alfv\'en waves (see lower panel of Fig. 1), which is significantly different from Braginskii (\cite{Braginskii1965}) and Forteza et al. (\cite{Forteza2007}). But, the slow magneto-acoustic waves have the same damping rate as that of Braginskii (\cite{Braginskii1965}). On the other hand, the damping rate of slow magneto-acoustic waves becomes different from the solution of Braginskii (\cite{Braginskii1965}) for higher plasma $\beta$. Fig. 7 shows the same as in the Fig. 6 but for $c_{sn}/v_A=0.5$. The damping rate of slow magneto-acoustic waves now begins to deviate from the solution of Braginskii for higher wave numbers. The behavior of fast magneto-acoustic waves remains the same.

\begin{figure}[t]
\vspace*{1mm}
\begin{center}
\includegraphics[width=9.5cm]{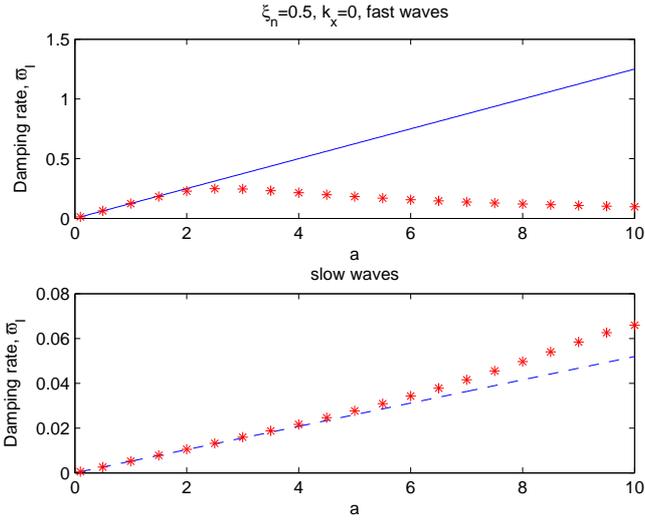}
\end{center}
\caption{The same as in Fig. 6 but for $c_{sn}/v_A=0.5$.}
\end{figure}

\section{Discussion}

Some parts of the solar atmosphere contain large number of neutral atoms: most of atoms are neutral at the photospheric level, but the ionization degree rapidly increases with height due to increased temperature. Solar prominences also contain neutral atoms. Neutral atoms may change the dynamics of plasma through collision with charged particles. For time scales longer than ion-neutral collision time, the partially ionized plasma can be considered as one fluid, because collisions between neutrals and charged particles lead to the rapid coupling of the two fluids. Then the equation of motion is written for the center of mass velocity, and the motion of species is considered as diffusion with a small velocity as compared to the velocity of center of mass. The corresponding collision terms appear in the equation of motion for the relative velocity (between ions and neutrals) and in the generalized Ohm's law. Neglecting the inertial term in the equation of motion for the relative velocity one simplifies the equations and traditional induction equation with Cowling conductivity is obtained (Braginskii \cite{Braginskii1965}, Khodachenko et al. \cite{Khodachenko2004}). The inertial terms (left-hand side terms in Eq. \ref{w1}) is smaller than the collision term (the last term in the same equation), but becomes comparable for time scales near ion-neutral collision time. Therefore, it can be neglected only for longer time-scales.

On the other hand, for the time scales less than ion-neutral collision time, the both fluids may behave independently and the single-fluid approximation is not valid any more. Then the two-fluid approximation, when ion-electron and neutral atom gases are treated as separate fluids, should be considered when one tries to model the processes in partially ionized plasmas.

The normal mode analysis of two-fluid partially ionized plasma shows that frequencies and damping rates of low-frequency MHD waves are in good coincidence with those found in the single-fluid approach. However, the waves with higher frequency than ion-neutral collision frequency have significantly different behavior. Alfv\'en and fast magneto-acoustic waves have maximal damping rates in particular frequency interval peaking, for example, at the frequency $\omega=2.5 \, \nu_{in}$, ($\nu_{in}$ is the ion-neutral collision frequency) for $\xi_n=0.5$ and at the frequency  $\omega=10 \, \nu_{in}$ for $\xi_n=0.1$. The damping rates are reduced for higher frequency part of wave spectrum (note that the damping rates are linearly increased in the single-fluid approach). Therefore, the statements concerning the damping of high-frequency Alfv\'en waves in the solar chromosphere due to ion-neutral collisions should be revised. Careful analysis is needed to study the damping of high frequency Alfv\'en waves for realistic height profile of ionization degree in the chromosphere.

Another important point concerning the Alfv\'en waves in partially ionized plasma is the cut-off wave-number, which appears in the single-fluid approach (Barc\'elo et al. \cite{Barcelo2010}). Barc\'elo et al. (\cite{Barcelo2010}) found that the Alfv\'en waves with larger wave numbers than the cut-off value are evanescent in partially ionized and resistive plasmas. However, our two-fluid analysis shows that there is no cut-off wave number due to ambipolar diffusion (see Fig.1, red asterisks). Therefore, the appearance of cut-off wave number in the single-fluid approach is the result of neglecting of inertial term in the equation of motion for relative velocity. It is possible that the cut-off wave number due to usual magnetic resistivity is caused by neglecting of electron inertia, therefore the cut-off may completely disappear in three-fluid approach. The cut-off wave numbers also appear for fast magneto-acoustic waves in partially ionized and resistive plasmas (Barc\'elo et al. \cite{Barcelo2010}). We suggest that this may also caused by the neglecting of inertial term. However, this point needs further study.

The two-fluid approach reveals two different slow magneto-acoustic modes when the slow wave time scale becomes shorter than ion-neutral collision time (Fig. 3). The different slow modes correspond to ion-electron and neutral  fluids. But, only one slow magneto-acoustic mode remains at lower frequency range as in usual single-fluid approach. This is easy to understand physically. When the wave frequency is lower than ion-neutral collision frequency, then the two fluids are coupled due to collisions and only one slow magneto-acoustic wave appears. The mode connected with the neutral fluid ("neutral" slow mode) has only imaginary frequency in this range of wave spectrum (not shown in the figure). This means that any slow wave type changes (density, pressure) in the neutral fluid is damped faster than wave period due to collisions with ions. The "neutral" slow magneto-acoustic wave has similar properties as ion magneto-acoustic waves.

It must be mentioned that the two-fluid approach of partially ionized plasma clarifies the uncertainty concerning the damping rate of slow magneto-acoustic waves found in the single-fluid approach. It was found that the normal mode analysis and energy consideration method (used by Braginskii \cite{Braginskii1965}) lead to different expressions for the damping rate of slow waves (Forteza et al. \cite{Forteza2007}). We found that the damping rate obtained in two-fluid approach is in good coincidence with the damping rate of Braginskii derived from the energy treatment (see lower panels of Fig. 5-6). Therefore, it seems that the discrepancy is again caused by neglecting the inertial terms in the equation of motion for the relative velocity in the single-fluid approach. Braginskii (\cite{Braginskii1965}) used the general energy method for the estimation of damping rates, and probably this is the reason why his results agree to those found in two-fluid approach.

Here we have considered only neutral hydrogen as a component of partially ionized plasma. However, other neutral atoms, for example neutral helium, may have important effects in MHD wave damping processes. Soler et al. (\cite{Soler2010}) made the first attempt to include the neutral helium in the single-fluid description of prominence plasma. They concluded that the neutral helium has not significant influence on the damping of MHD waves. However, two-fluid approach may give some more details about the effects of neutral Helium on MHD waves, therefore it is important to study this point in the future.

\section{Conclusions}

\begin{enumerate}
      \item Frequencies and damping rates of low frequency MHD waves in the two-fluid description are similar to those obtained in the single-fluid approach. But high-frequency waves (with higher frequency than the ion-neutral collision frequency) have completely different behavior.
     \item Alfv\'en and fast magneto-acoustic waves have maximal damping rates at some frequency interval peaking at particular frequency. The peak frequency is $2.5 \, \nu_{in}$, where $\nu_{in}$ is the ion-neutral collision frequency, for 50$\%$ of neutral hydrogen. For 10$\%$ of neutral hydrogen, the peak frequency is shifted to$10 \, \nu_{in}$. The damping rate is reduced for higher frequencies, therefore the damping of high-frequency Alfv\'en waves in the solar chromosphere with realistic height profile of ionization degree needs to be revised in future.
      \item There are two types of slow magneto-acoustic waves in the high-frequency part of wave spectrum: one connected with the ion-electron fluid and another with the fluid of neutrals.
      \item There is no cut-off frequency of Alfv\'en waves due to ambipolar diffusion. The cut-off frequency found in the single-fluid approach is caused by neglecting the inertial terms in the momentum equation of relative velocity.
      \item The damping rate of slow magneto-acoustic waves is similar to Braginksii (\cite{Braginskii1965}) in low plasma $\beta$ approximation. The deviation from the Braginskii formula found by normal mode analysis in single-fluid approach (Forteza et al. \cite{Forteza2007}) is probably caused by neglecting the inertial terms.
   \end{enumerate}

%

\begin{acknowledgements}
The work was supported by the Austrian Fond zur F\"orderung
der wissenschaftlichen Forschung (project P21197-N16). T.V.Z. also acknowledges
financial support from the Georgian National Science Foundation (under grant GNSF/ST09/4-310).
\end{acknowledgements}

\appendix
\section{Single-fluid equations}

We use the total velocity (i.e. velocity of center of mass)
\begin{equation}\label{V1}
\vec V= {{\rho_{i}\vec V_i+\rho_{n}\vec V_n}\over
{\rho_{i}+\rho_{n}}},\,\
\end{equation}
relative velocity
\begin{equation}\label{w}
\vec w= \vec V_i - \vec V_n.
\end{equation}
and total density
\begin{equation}\label{rho}
\rho=\rho_{i}+\rho_{n}.
\end{equation}

Eqs. (\ref{ni2})-(\ref{pn2}) and (\ref{B2}) lead to
the system:
\begin{equation}\label{n1}
{{\partial \rho}\over {\partial t}}+\nabla \cdot (\rho \vec V)=0,
\end{equation}
\begin{equation}\label{V1}
\rho{{\partial \vec V}\over {\partial t}}+\rho({\vec V}\cdot
\nabla)\vec V=-\nabla p+{1\over c}\vec j \times \vec B - \nabla
\cdot (\xi_i \xi_n \rho \vec w \vec w),
\end{equation}
$$
{{\partial \vec w}\over {\partial t}}+({\vec V}\cdot \nabla)\vec
w+({\vec w}\cdot \nabla)\vec V+\xi_n({\vec w}\cdot \nabla)\vec
w-({\vec w}\cdot \nabla)\xi_i \vec w=
$$
\begin{equation}\label{w1}
-\left ({{\nabla p_{ie}}\over
{\rho \xi_i}}-{{\nabla p_n}\over {\rho \xi_n}}\right )+{1\over {c
\rho \xi_i}}\vec j \times \vec B+{{\alpha_{en}}\over {e n_e \rho
\xi_i \xi_n}}\vec j -{{\alpha_{n}}\over {\rho \xi_i \xi_n}}\vec w,
\end{equation}
$$
{{\partial p}\over {\partial t}}+ ({\vec V}\cdot \nabla)p + \gamma p
\nabla \cdot \vec V -\xi_i({\vec w}\cdot \nabla)p -\gamma p \nabla
\cdot (\xi_i \vec w)+
$$
$$
({\vec w}\cdot \nabla)p_{ie}+\gamma p_{ie}
\nabla \cdot \vec w=(\gamma -1) {{\alpha_{ei}+\alpha_{en}}\over {e^2
n^2_e}}j^2 +(\gamma -1)\alpha_{n}w^2-
$$
\begin{equation}\label{p1}
(\gamma -1){{2 \alpha_{en}}\over {e n_e}}\vec j \vec w +{1\over {e
n_e}}({\vec j}\cdot \nabla)p_{e}+ \gamma p_{e}\nabla \cdot {{\vec
j}\over {e n_e}}-(\gamma -1)\nabla \cdot ({\vec q_i}+{\vec
q_e}+{\vec q_n}),
\end{equation}
$$
{{\partial \vec B}\over {\partial t}}={\nabla \times}(\vec V \times
\vec B)+ {\nabla \times}\left ({{c \nabla p_e}\over {e n_e}}\right
)-{\nabla \times}\left (\eta \nabla \times \vec B\right ) - {\nabla
\times}\left ({{\vec j\times \vec B}\over {en_e}}\right )+
$$
\begin{equation}\label{B1}
{\nabla
\times}\left ({{c \alpha_{en}\vec w}\over {e n_e}}\right )+ {\nabla
\times}\left (\xi_n\vec w \times \vec B\right ),
\end{equation}
where $p=p_e+p_i+p_n$, $\xi_i=\rho_i/\rho$,
$\xi_n=\rho_n/\rho$ and $\alpha_{n}=\alpha_{in}+\alpha_{en}$.

The Ohm's law is now
\begin{equation}\label{om1}
\vec E + {1\over c}\vec V \times \vec B+ {{1}\over {e n_e}}\nabla
p_e={{\alpha_{ei}+\alpha_{en}}\over {e^2 n^2_e}}\vec
j-{{\alpha_{en}}\over {e n_e}}\vec w +{1\over {cen_e}}\vec j\times
\vec B -{\xi_n\over c}\vec w \times \vec B.
\end{equation}

Neglecting the inertia terms i.e. all left hand side terms in Eq. (\ref{w1}) we have
$$
\vec w= - {{\vec G}\over \alpha_n} + {{\xi_n}\over {c \alpha_n}}\vec
j\times\vec B + {{\alpha_{en}}\over {e n_e \alpha_n}}\vec j.
$$
Then the induction equation takes the form
$$
{{\partial \vec B}\over {\partial t}}={\nabla \times}(\vec V \times
\vec B)+ {c\over e}{\nabla \times}\left ({{\nabla p_e -\epsilon \vec
G}\over {n_e}}\right )-{\nabla \times}\left (\eta_T \nabla \times
\vec B\right ) -
$$
$$
{c\over {4 \pi e}}{\nabla \times}\left
({{1-2\epsilon \xi_n}\over {n_e}}(\nabla \times \vec B)\times \vec
B\right ) - {\nabla \times}\left ({{\xi_n}\over {\alpha_n}}\vec G
\times \vec B\right )+
$$
\begin{equation}\label{B1}
{\nabla \times}\left ({{\xi^2_n}\over {4 \pi \alpha_n}}((\nabla
\times \vec B)\times \vec B)\times \vec B\right ),
\end{equation}
where $\epsilon=\alpha_{en}/\alpha_n$, $\vec G=\xi_n\nabla
p_{ei}-\xi_i\nabla p_{n}$ and
\begin{equation}\label{etaT}
\eta_T={{c^2}\over {4\pi \sigma}}={{c^2}\over {4\pi e^2
n^2_e}}(\alpha_{ei}+\alpha_{en}-{{\alpha^2_{en}}\over \alpha_n}).
\end{equation}

These equations are traditionally used for the description of partially ionized plasmas.


\end{document}